# Orbital-selective charge-density wave in TaTe$_4$

R. Z. Xu[1], X. Du[1], J. S. Zhou[1], X. Gu[1], Q. Q. Zhang[1], Y. D. Li[1], W. X. Zhao[1], F. W. Zheng[2], M. Arita[3], K. Shimada[3], T. K. Kim[4], C. Cacho[4], Y. F. Guo[5,6], Z. K. Liu[5,6], Y. L. Chen[1,5,6,7], L. X. Yang[1,8,9]

[1]*State Key Laboratory of Low Dimensional Quantum Physics, Department of Physics, Tsinghua University, Beijing 100084, China*

[2]*Institute of Applied Physics and Computational Mathematics, Beijing 100088, China*

[3]*Hiroshima Synchrotron Radiation Center, Hiroshima University, Higashi-Hiroshima 739-0046, Japan*

[4]*Diamond Light Source, Harwell Science & Innovation Campus, Didcot, Oxfordshire – OX11 0QX, U.K.*

[5]*School of Physical Science and Technology, ShanghaiTech University and CAS-Shanghai Science Research Center, Shanghai 201210, China.*

[6]*ShanghaiTech Laboratory for Topological Physics, Shanghai 200031, China.*

[7]*Clarendon Laboratory, Department of Physics, University of Oxford, Oxford OX1 3PU, U.K.*

[8]*Frontier Science Center for Quantum Information, Beijing 100084, China*

[9]*Collaborative Innovation Center of Quantum Matter, Beijing 100871, China*

**TaTe$_4$, a metallic charge-density wave (CDW) material discovered decades ago, has attracted renewed attention due to its rich interesting properties such as pressure-induced superconductivity and candidate non-trivial topological phase. Here, using high-resolution angle-resolved photoemission spectroscopy and *ab-initio* calculation, we systematically investigate the electronic structure of TaTe$_4$. At 26 K, we observe a CDW gap as large as 290 meV, which persists up to 500 K. The CDW-modulated band structure shows a complex reconstruction that closely correlates with the lattice distortion. Inside the CDW gap, there exist highly dispersive energy bands contributing to the remnant Fermi surface and metallic behavior in the CDW state. Interestingly, our *ab-initio* calculation reveals that the large CDW gap mainly opens in the electronic states with out-of-plane orbital components, while the in-gap metallic states originate from in-plane orbitals, suggesting an orbital texture that couples with the CDW order. Our results shed light on the interplay between electron, lattice, and orbital in quasi-one-dimensional CDW materials.**

## INTRODUCTION

Charge density waves (CDWs), long-range order of electron density on the positive background of periodic lattice distortion (PLD), have been discovered and extensively studied for decades[1-3]. While CDWs have been widely observed and suggested to be closely related to many intriguing emergent phenomena, such as high-temperature superconductors in cuprates[4], heavy fermion behavior[5], and topological quantum states[6-8], the mechanism of CDW transition is still under debate and seems to be materials-dependent[9-13]. As originally proposed by Peierls, the perfect Fermi surface (FS) nesting in one-dimensional metallic systems induces a divergence in the electronic response function and enhances the electronic screening of the phonons with the momentum of twice the electron Fermi momentum, which facilitates the formation of the PLD and CDW. In real materials, however, this picture has been challenged since the FS nesting is usually imperfect and incapable to drive the lattice distortion[12]. In some systems, such as the Luttinger liquid materials,

CDW can also emerge with strongly suppressed electronic states near the Fermi level ($E_F$)[9,14-16]. Alternatively, a scenario of momentum-dependent electron-phonon coupling was proposed to play a vital role in the formation of CDWs[12,17].

In addition, the orbital degree of freedom is also involved in the CDW ordering in many two-dimensional CDW materials[18-25]. On the one hand, complex orbital texture has been shown to intertwine with the CDWs in different materials[22,26]. On the other hand, the orbital hybridization and ordering can strongly modulate the electronic states near $E_F$, thus affecting or driving the CDW formation. For example, in the monolayer $TiTe_2$, the selective hybridization between orbitals can greatly save the total energy of the system and favor the CDW transition[21]. Recently, orbital hybridization is likewise shown to be able to drive the CDW transition in the kagome material $CsV_3Sb_5$[27,28]. It is therefore of paramount importance to investigate the interplay between electron, orbital, and lattice in the CDW ordered states for a comprehensive picture of CDW transitions.

Transition metal tetrachalcogenide $TaTe_4$ is a prototypical quasi-one-dimensional (Q1D) CDW material discovered in the 1960s[29-35]. Recently, it attracts renewed research interests due to its rich intriguing properties such as superconductivity under high pressure[36], nontrivial topological properties[34,37,38], and large anisotropic magnetoresistance[37,38]. After extensive studies, however, there are still mysteries in the CDW properties of $TaTe_4$. Firstly, $TaTe_4$ shows metallic behavior in the CDW state, despite a large CDW gap of about 300 meV observed by optical conductivity and angle-resolved photoemission spectroscopy (ARPES) measurements[31]. Secondly, although $TaTe_4$ exhibits a Q1D crystal structure, no Q1D band dispersion was directly observed in previous ARPES measurements[31]. Thirdly, the mechanism of the CDW transition in $TaTe_4$ is still elusive: while the FS nesting is excluded by recent calculations of the electronic structure, it is still under debate whether the softened phonon modes originate from strong electron-phonon coupling or structural optimization[35,39]. Moreover, the surface CDW modulation is sensitive to the external magnetic field, suggesting a complex interplay between different degrees of freedom in the CDW ordering[32].

In this work, we study the electronic structure of TaTe$_4$ using high-resolution ARPES. At low temperatures, we observe a CDW gap of about 290 meV, much larger than the mean-field estimation of 3.5 $k_B T_c$, where $T_c$ is the CDW transition temperature of TaTe$_4$. Inside the CDW gap, we observe dispersive bands forming Q1D FS sheets and accounting for the metallic behavior in the CDW state. Our *ab-initio* calculation confirms the irrelevance of FS nesting in the CDW transition, consistent with previous results. Moreover, the large CDW gap selectively opens in the electronic states of out-of-*ab*-plane orbital components, while the in-gap band dispersions of in-*ab*-plane orbital components show weak reconstruction, suggesting an intriguing orbital-selective CDW formation in TaTe$_4$. Our results provide important insights into the interplay between electron, lattice, and orbital degrees of freedom in the CDW state of Q1D TaTe$_4$.

**RESULTS**

**Basic properties of TaTe$_4$**

TaTe$_4$ crystallizes in a tetragonal structure with *P4/mcc* space group in the normal state[39]. Each Ta atom is coordinated with eight Te atoms, forming Q1D chains of face-sharing square antiprisms along the *c* direction, as schematically shown in Fig. 1a. With decreasing temperature, the system first enters a $\sqrt{2}a \times \sqrt{2}a \times 3c$ commensurate CDW (CCDW) state, then a $2a \times 2a \times 3c$ CCDW′ state below $T'_{CCDW} \approx 450$ K (Fig. 1b)[30,40,41]. The CCDW′ transition is manifested in the magnetic susceptibility measurement in Fig. 1c, where we observe an anomaly in the slope of the $\chi$-$T$ curve near 450 K. Figure 1d shows the Raman scattering data at selected temperatures. With increasing temperature, most of the Raman peaks shift toward lower energies. We noted that the peak at about 58 cm$^{-1}$ (marked by the black arrow) shows a minor shift but a significant reduction of its intensity (about 80%), which may be related to the CCDW′ transition (Supplementary Note 1).

Figure 1e and 1f show the *ab-initio* calculations of the band structure of TaTe$_4$ in the normal state without and with the spin-orbit coupling (SOC) effect. SOC exerts a strong influence on the electronic structure by opening large gaps at the band crossings along Γ-*X*-*M*-Γ as shown in Fig.

1f. The Fermi momenta of the energy bands, however, is far from the CDW wavevector, as exemplified by the black arrow along $\Gamma Z$ in the inset of Fig. 1f. Even after the calculated chemical potential is aligned to the experimental value (Supplementary Note 4), the theoretical fermi momenta is still different from the CDW wave vector, confirming the irrelevance of the FS nesting in the CDW transition[12].

Figure 2 shows the calculated and ARPES-measured FS of TaTe$_4$. In the normal state, we reveal three FS sheets: a warped Q1D open FS (blue sheets in Fig. 2a) and two 3D closed pockets around the $M$ and $A$ points (green sheets in Fig. 2a). To better compare with the experiments, Fig. 2b shows the overlaid FS in the $\Gamma ZRX$ and $XRAM$ planes. In the $2 \times 2 \times 3$ CCDW′ state, the FS shows a drastic reconstruction. We observe four FS sheets in the Brillouin zone of the CCDW′ state (Fig. 2c). The green open FS sheet is nearly dispersionless along $k_z$, which may be important for the intriguing magnetotransport properties of TaTe$_4$[37,38]. Fig. 2d shows the 2D plot of the FS in the CCDW′ state in the $\gamma zrx$ plane.

Since the sample is cleaved along the (010) plane as shown in Fig. 1a, ARPES measurement is conducted in the $k_x$-$k_z$ plane. Figure 2e presents the measured FS with a large electron pocket at the $\bar{X}$ point and quasi-1D FS sheets along $k_x$. Both the Fermi surface sheets are captured by the calculation despite the differences in the fine details. It is worth noting that the band structure along $\Gamma X$ can also be measured by photon-energy-dependent experiments due to the tetragonal crystal structure. Figure 2f presents the FS measured along $k_y$. We observe a clear reconstruction of the Q1D FS sheets along $k_z$ (due to the weak intensity at $E_F$, we show the data at 150 meV below $E_F$) within the expectation of the 3-fold band folding along $k_z$ (Supplementary Note 5). The folding of the in-gap states induces eye-shape features in the $k_x$-$k_z$ map, which can be better visualized in the constant-energy-contours at higher binding energies (Supplementary Note 3)[42].

**Band reconstruction with the CDW formation**

Figure 3a shows the measured band dispersions along $\bar{\Gamma}\bar{Z}$. Consistent with the previous measurement[31], we observe a large gap of about 290 meV and multiple dispersive bands in the high binding energies (also see Supplementary Notes 6 and 7). Besides, there exist noticeable features with weak intensity in the large CDW gap. In the zoom-in plot of ARPES spectra in the CDW gap region (Fig. 3b), we reveal highly dispersive bands crossing $E_F$, which contribute to the FS in the CCDW′ state and account for the metallic transport behavior of the system at low temperatures[31]. To investigate the interplay between the electronic reconstruction and lattice distortion, we calculate the electronic structure of TaTe$_4$ along ΓZ with different PLD magnitudes as shown in Fig. 3c-3f. With the development of the lattice distortion, the folded bands hybridize with the original bands and the CDW gap gradually increases. At 100% PLD distortion level (corresponding to experimental lattice distortion), the CDW gap is as large as 300 meV, in good agreement with our experiment. Moreover, the in-gap states are well captured by the calculation (Fig. 3f). By comparing the measured and calculated band structures, we reveal clear evidence for the band reconstruction due to CDW superstructure along the *c* direction (Supplementary Note 2).

**Orbital-selective CDW gap formation**

We notice that the magnitude of the CDW gap varies in different bands. For example, in Fig. 3c, the α band (highlighted by yellow thick lines in Fig. 3d-f) is much more sensitive to the lattice distortion than the β band (highlighted by green thick lines). To understand this difference, we calculate the orbital-projected band structure of TaTe$_4$ in Fig. 4. As shown in Fig. 4a and 4b, along the ΓZ direction, the α (β) band in the normal state mainly consists of out-of-*ab*-plane (in-*ab*-plane) Ta $5d_{z^2}$ ($5d_{x^2-y^2}$) orbital hybridized with Te $5p_z$ ($5p_{x/y}$) orbital. In the CCDW′ state, the $d_{z^2}$ orbital components are strongly affected by opening a large CDW gap, while the $d_{x^2-y^2}$ orbital components show weak change. Along other momentum directions, the $d_{x^2-y^2}$ orbitals are likewise highly dispersive and contribute to band dispersions crossing $E_F$ (Fig. 4c and 4d), while the global orbital-selective CDW gap induces narrow bands (bandwidth ~200 meV) with $d_{z^2}$ orbitals. Such an orbital-dependent reconstruction of the electronic structure with CDW ordering resembles the orbital texture in the CCDW state of 1T-TaSe$_2$[22], suggesting an intriguing interplay between charge, lattice, and orbital degrees of freedom in TaTe$_4$.

## DISCUSSIONS

Our experiment and calculation suggest that the FS nesting is irrelevant in the CDW transition of TaTe$_4$. Firstly, the Fermi momenta of the calculated and experimental band structure deviate from the CDW wavevector along $k_z$. Secondly, we observe a reduced CDW gap of $\frac{2\Delta}{k_B T'_{CCDW}} \approx 15$, which is much larger than the mean-field estimation, evincing a strong-coupling nature of the CDW (even with the highest reported CCDW transition temperature of 650 K[43], the reduced gap is still much larger than 3.5). Correspondingly, the distortion of Ta atoms is about 6.6% of the lattice constant[44], similar to that in 1T-TaS$_2$, a well-known strong-coupling CDW material[45]. Thirdly, previous calculations of the phonon spectrum revealed a broad phonon softening, which was understood by the strong electron-phonon interaction instead of Fermi surface nesting[39]. The phonon frequencies decrease in a broad momentum range and drop to minus values around $\vec{q}_{CDW}$ [39] similar to the situation in NbSe$_2$[46] and monolayer VTe$_2$[17] where the CDW transition is explained by a momentum-dependent EPC model[17,46,47].

On the other hand, while CDW is an electron-phonon intertwined state, other degrees of freedom such as orbital and spin can be deeply involved by modulating the electronic states near $E_F$[23,48-50]. In TaTe$_4$, due to the Q1D crystal structure, the lattice distortion is highly anisotropic. In the CDW state at low temperatures, the Ta-Te bonds expand by about 2.8% and the Ta-Ta distance remains unchanged. By contrast, the displacement of Ta atoms is as large as 6.6% of $c$ along the $z$-axis[44]. This anisotropic lattice distortion provides a structural basis for the strong interaction between the $d_{z^2}$ orbital-derived electronic states. The interplay between electron, lattice, and orbital results in the intriguing orbital-selective CDW state with metallic in-gap states, which provides an interesting platform to explore the orbital-selective transport properties or orbitronics.

In summary, our results show that the metallic properties of CDW ordering in TaTe$_4$ originate from dispersive in-gap states. We reveal an orbital-selective formation of the CDW gap, suggesting a complex interplay between electron, lattice, and orbital degrees of freedom in the CDW transition of TaTe$_4$.

**METHODS**

**Sample preparation**

High-quality TaTe$_4$ single crystals were synthesized by the self-flux method using tellurium flux. Tantalum (99.9%, Aladdin Chemicals) and tellurium (99.999%, Aladdin Chemicals) powders were mixed in the molar ratio of Ta: Te = 1: 30 and placed into an alumina crucible. Under a vacuum lower than $10^{-4}$ Pa, the crucible is sealed into a quartz tube and heated slowly up to 1000 °C within 15 hours. The temperature was kept at 1000 °C for around 20 hours so that the two substances fully react. After cooling the system uniformly down to 550 °C, the quartz tube was quickly transferred to a centrifuge to separate the excess tellurium. We examine the crystalline phase and sample qualities by Bruker D8 VENTURE single crystal diffractometer utilizing Mo K$_\alpha$1 radiation with a wavelength of 0.7093 Å.

**ARPES measurements**

High-resolution ARPES experiments were performed at beamline 9A of Hiroshima Synchrotron Radiation Center (HSRC) and beamline I05 of Diamond Light Source (DLS)[51] using Scienta R4000 analyzers. ARPES measurements above 400 K were conducted at Peking University using a Scienta DA30L analyzer and a helium discharge lamp. The convolved energy and angular resolutions were 15 meV and 0.2° respectively. The samples were cleaved *in situ* and measured under ultrahigh vacuum below $1 \times 10^{-10}$ mbar.

**Band structure calculations**

The distortion-dependent band structures were calculated using Vienna Ab initio Software Package (vasp)[52]. The atom core electrons were described by the projector augmented wave (PAW) method[53,54]. The Perdew-Burke-Ernzerhof (PBE) functional[55] was used to treat the electronic exchange correlation. The plane-wave energy cut-off was set to 350 eV. The reciprocal space was sampled with a Monkhorst-Pack grid $8 \times 8 \times 8$ for the normal-state unit-cell and a grid $3 \times 3 \times 2$ for $2a \times 2a \times 3c$ supercell. The cut-off energy and Monkhorst-pack grids were converged to produce reliable electronic structures. The atomic structure was relaxed until the force on each atom is smaller than 0.01 eV/A. We have used the van der Waals correction[56] and spin-orbit coupling correction for the density-functional-theory calculations.

The orbital-projected band structures were calculated using QUANTUM ESPRESSO package[57] with a plane wave basis. The exchange-correlation energies were considered under Perdew-Burke-Ernzerhof (PBE) type generalized gradient approximation (GGA)[55]. Both cases of excluding and including spin-orbit coupling were considered. The cutoff energy for the plane-wave basis was set to 400 eV. A Γ-centered Monkhorst-Pack *k*-point mesh of $7 \times 7 \times 7$ was adopted for a self-consistent charge density.

## DATA AVAILABILITY

The data supporting the conclusions in this work are available from the corresponding authors.

## ACKNOWLEDGMENTS

This work was supported by the National Natural Science Foundation of China (Grant No. 12274251) and the National Key R&D program of China (Grants No. 2022YFA1403201 and 2022YFA1403100). L.X.Y. acknowledges the support from the Tsinghua University Initiative Scientific Research Program and the Fund of Science and Technology on Surface Physics and Chemistry Laboratory (No. XKFZ202102). Y.L.C. acknowledges the support from the Oxford-ShanghaiTech collaboration project and the Shanghai Municipal Science and Technology Major Project (grant 2018SHZDZX02). Y.F.G. acknowledges the the Double First-Class Initiative Fund of ShanghaiTech University. We thank Diamond Light Source for access to beamline I05 (Proposal number SI20683-1) that contributed to the results presented here. ARPES experiments at the Hiroshima Synchrotron Radiation Center (HiSOR), Hiroshima University were done under Proposal No. 18BG051. We thank N-BARD, Hiroshima University, for supplying the liquid helium.

**COMPETING INTERESTS**

The authors declare no Competing Financial or Non-Financial Interests.

**AUTHOR CONTRIBUTIONS**

L.X.Y. and Y.L.C. conceived and supervised the experiments. R.Z.X. and X.D. contributed equally to this paper. R.Z.X. carried out the ARPES experiments and corresponding analysis with the assistance of X.D., J.S.Z., X.G., Q.Q.Z., Y.D.L., W.X.Z., M.A., K.S., T.K.K., C.C., Z.K.L. X.D. performed the *ab-initio* calculations with the assistance of F.W.Z. Y.F.G provided high-quality single crystal samples. R.Z.X. wrote the first draft of the paper. L.X.Y., Z.K.L. and Y.L.C. contributed to the revision of the manuscript. R.Z.X and X.D. contributed to this work equally. All authors contributed to the scientific planning and discussions.

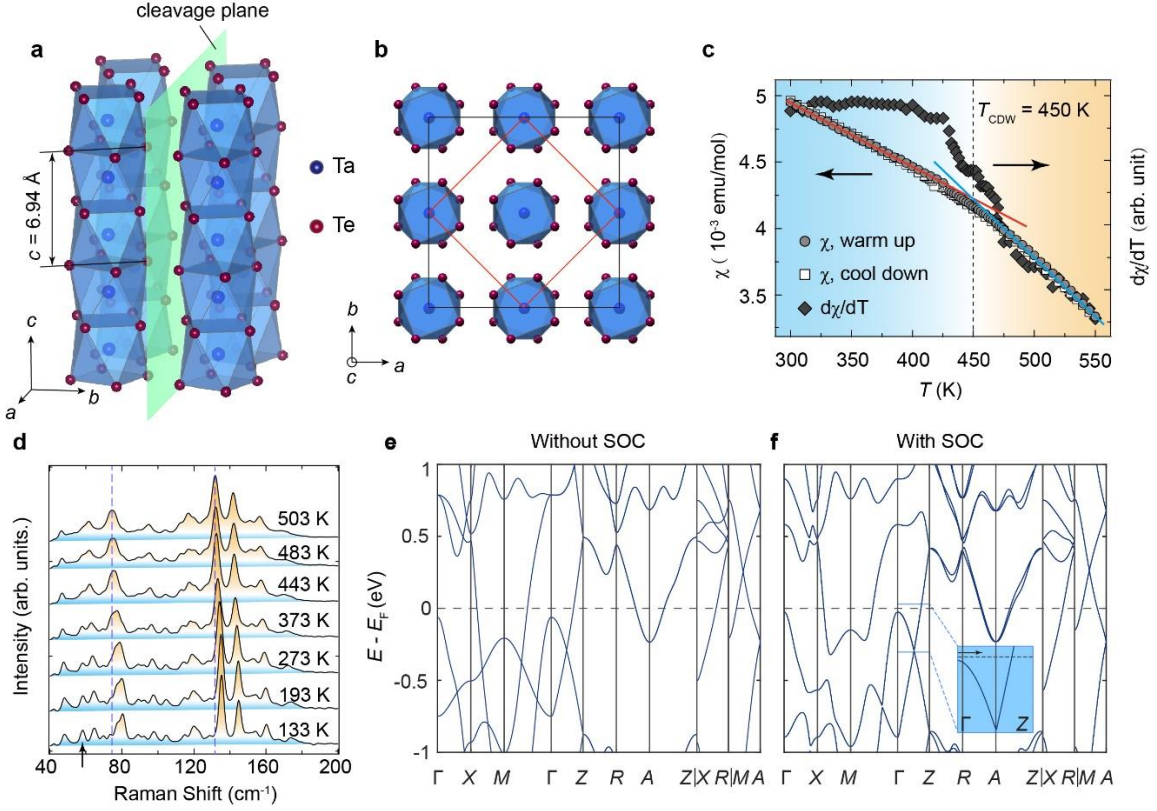

**Fig. 1. Basic properties of TaTe$_4$. a-b** Crystal structure of TaTe$_4$ in the front view (**a**) and top view (**b**). The green plane indicates the cleavage plane. Black and red lines indicate the superlattices of the $2a \times 2a \times 3c$ CCDW′ and $\sqrt{2}a \times \sqrt{2}a \times 3c$ CCDW states. **c** Magnetic susceptibility of TaTe$_4$ as a function of temperature. The red and blue lines are guides to eyes for the change of the slope near $T'_{\text{CCDW}} = 450$ K. **d** Raman scattering spectra at selected temperatures from 130 K to 503 K. **e-f** Calculated band structure of TaTe$_4$ in the normal state without (**e**) and with (**f**) spin-orbit coupling (SOC). The inset shows the band structure along ΓZ near the Fermi level ($E_F$) with the black arrow indicating the CCDW′ wavevector.

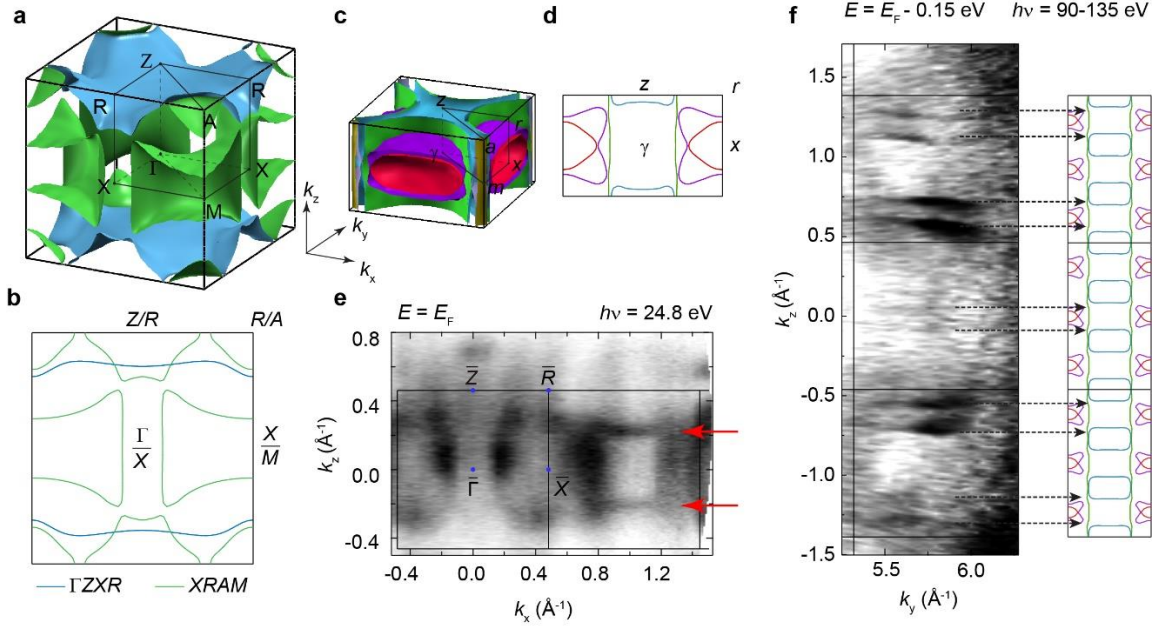

**Fig. 2. Calculated and experimental Fermi surface (FS) of TaTe$_4$. a** Calculated FS in the normal state in the bulk Brillouin zone (BZ) (**a**). **b** Overlaid FS contours in the Γ*ZRX* (blue curves) and *XRAM* (green curves) planes. **c** Calculated FS in the CCDW' state. **d** FS contour in the CCDW' state in the γ*zrx* plane. The uppercase and lowercase letters mark the high symmetry points of the BZ in the normal states and CCDW' states respectively. **e** FS measured using linear-horizontally polarized photons at 24.8 eV at 40 K. **f** Constant-energy-contour at 0.15 eV below $E_F$ measured by photon-energy-dependent experiments using linear-horizontally (LH) polarized photons at 40 K. The photon energy ranges from 90 to 135 eV. The calculated result at 0.21 eV below $E_F$ considering the chemical potential difference between the experiment and calculation (Supplementary Note 4) is shown side-by-side for comparison.

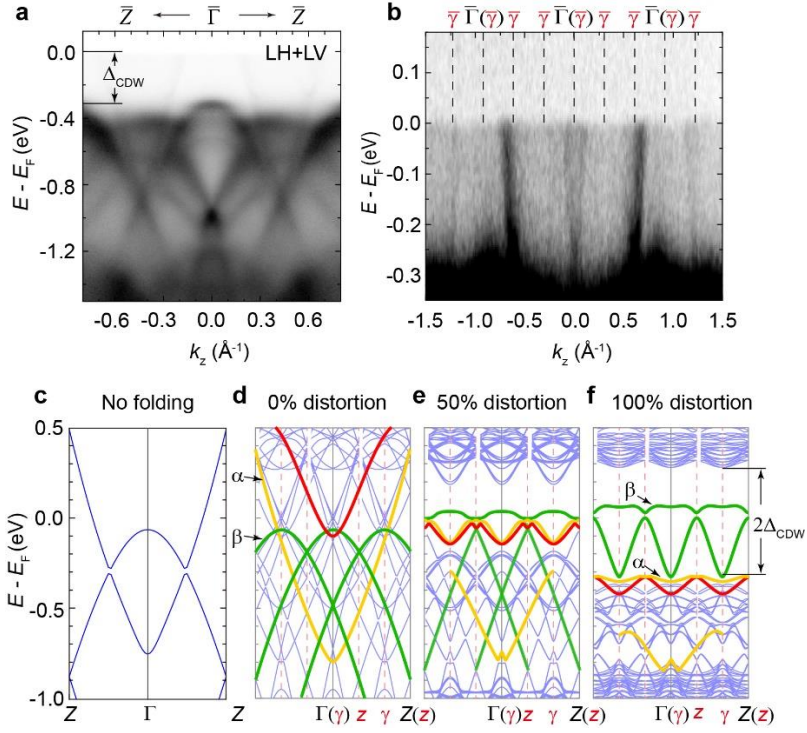

**Fig. 3. Evolution of the band structure of TaTe$_4$ with periodic lattice distortion (PLD). a** Band dispersions along the $\bar{\Gamma}\bar{Z}$ direction measured at 40 K. The ARPES spectra measured by LH and linear-vertically (LV) polarized photons are summed together. **b** Zoom-in plot of the band structure in the CDW gap. **c** Calculated electronic structure of TaTe$_4$ in the normal state. **d-f** Evolution of the calculated band structure along the $\Gamma Z$ direction with increasing PLD magnitudes. The yellow and green lines in **d** are the bands along $\Gamma Z$. The red line in **d** is the band along $XR$ which was folded to the $\Gamma Z$ direction. Data in **a** and **b** were collected using 24.8 eV and 100 eV photons respectively.

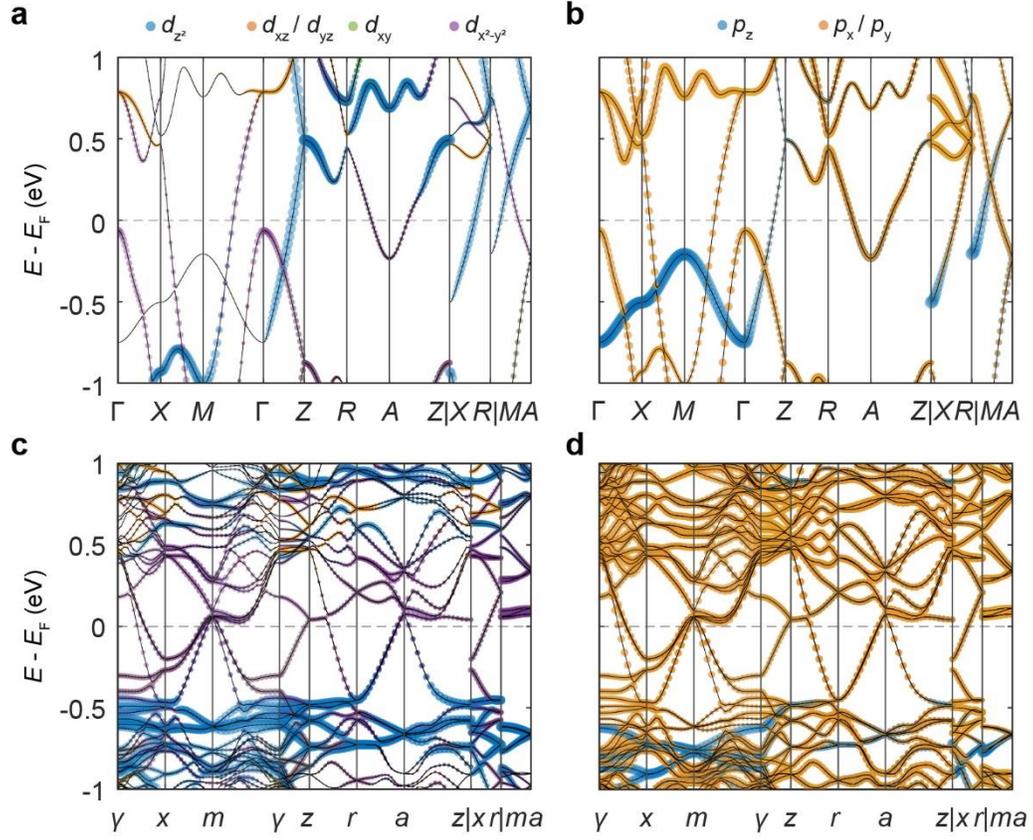

**Fig. 4. Orbital-projected calculations of the band structure of TaTe$_4$. a- b** Contribution from Ta *d*-orbitals (**a**) and Te *p*-orbitals (**b**) in the normal state. **c-d** Same as **a-b** but in the CCDW′ state. The *xyz* framework corresponds to the lattice coordinate *abc* respectively.

**Reference**

1. Grüner, G. & Zettl, A. Charge density wave conduction: A novel collective transport phenomenon in solids. *Phys. Rep.* **119**, 117-232 (1985).
2. Grüner, G. The dynamics of charge-density waves. *Rev. Mod. Phys.* **60**, 1129 (1988).
3. Gor'kov, L. P. & Grüner, G. *Charge density waves in solids*. (Elsevier, 2012).
4. Chang, J. *et al.* Direct observation of competition between superconductivity and charge density wave order in $YBa_2Cu_3O_{6.67}$. *Nat. Phys.* **8**, 871-876 (2012).
5. Neilson, J. R. *et al.* Mixed-valence-driven heavy-fermion behavior and superconductivity in $KNi_2Se_2$. *Phys. Rev. B* **86**, 054512 (2012).
6. Shi, W. *et al.* Publisher Correction: A charge-density-wave topological semimetal. *Nat. Phys.* **17**, 284 (2021).
7. Cho, S. *et al.* Emergence of New van Hove Singularities in the Charge Density Wave State of a Topological Kagome Metal $RbV_3Sb_5$. *Phys. Rev. Lett.* **127**, 236401 (2021).
8. Qin, F. *et al.* Theory for the charge-density-wave mechanism of 3D quantum Hall effect. *Phys. Rev. Lett.* **125**, 206601 (2020).
9. Kang, L. *et al.* Band-selective Holstein polaron in Luttinger liquid material $A_{0.3}MoO_3$ (A= K, Rb). *Nat. Commun.* **12**, 6183 (2021).
10. Schäfer, J. *et al.* Unusual spectral behavior of charge-density waves with imperfect nesting in a quasi-one-dimensional metal. *Phys. Rev. Lett.* **91**, 066401 (2003).
11. Perfetti, L. *et al.* Mobile small polarons and the Peierls transition in the quasi-one-dimensional conductor $K_{0.3}MoO_3$. *Phys. Rev. B* **66**, 075107 (2002).
12. Johannes, M. & Mazin, I. Fermi surface nesting and the origin of charge density waves in metals. *Phys. Rev. B* **77**, 165135 (2008).
13. Shen, D. *et al.* Novel mechanism of a charge density wave in a transition metal dichalcogenide. *Phys. Rev. Lett.* **99**, 216404 (2007).
14. Du, X. *et al.* Crossed Luttinger liquid hidden in a quasi-two-dimensional material. *Nat. Phys.* **19**, 40-45 (2023).
15. Ejima, S., Hager, G. & Fehske, H. Quantum phase transition in a 1D transport model with Boson-affected hopping: luttinger liquid versus charge-density-wave behavior. *Phys. Rev. Lett.* **102**, 106404 (2009).
16. Hohenadler, M., Wellein, G., Bishop, A., Alvermann, A. & Fehske, H. Spectral signatures of the Luttinger liquid to the charge-density-wave transition. *Phys. Rev. B* **73**, 245120 (2006).15